\documentclass[aps, a4paper,superscriptaddress, nofootinbib, showpacs, twocolumn]{revtex4} 
\usepackage{amsmath,amssymb,amsfonts,dsfont}
\usepackage{booktabs,tabulary}
\usepackage{epsfig}
\usepackage{graphicx}
\usepackage{color}
\usepackage{placeins}
\usepackage{multirow}

\newcommand{\be}{\begin{equation}}
\newcommand{\ee}{\end{equation}}
\newcommand{\beq}{\begin{equation}}
\newcommand{\eeq}{\end{equation}}
\newcommand{\bea}{\begin{eqnarray}}
\newcommand{\eea}{\end{eqnarray}}

\newcommand{\gev}{\, \text{GeV}}
\newcommand{\mev}{\, \text{MeV}}

\renewcommand{\Im}{\text{Im}\,}
\newcommand{\diff}{\text{d}}

\let\ifcomments\iftrue
\def\commentsoff{\global\let\ifcomments\iffalse}
\let\commentsize\small
\def\tinycomments{\global\let\commentsize\footnotesize}

\usepackage[normalem]{ulem}

\begin{document}

\begin{flushleft}
IFT-UAM/CSIC-16-010, FTUAM-16-4, MITP/16-014
\end{flushleft}

~\vspace{0.5cm}

\title{On the uncertainty estimates of the $\sigma$-pole determination by Pad\'e approximants}

\author{Irinel Caprini} 
 \affiliation{Horia Hulubei National Institute for Physics and Nuclear Engineering,
   P.O.B. MG-6, 077125 Bucharest-Magurele, Romania}

\author{Pere Masjuan} 
 \affiliation{PRISMA Cluster of Excellence, Institut f\"ur Kernphysik, Johannes Gutenberg-Universit\"at, D-55099 Mainz, Germany }

\author{Jacobo Ruiz de Elvira}
\affiliation{Helmholtz-Institut f\"ur Strahlen- und Kernphysik, Universit\"at Bonn, D-53115
Bonn, Germany}

\author{Juan Jos\'e Sanz-Cillero}
\affiliation{Departamento de F\'isica Te\'orica and Instituto de F\'isica Te\'orica, IFT-UAM/CSIC
       Universidad Aut\'onoma de Madrid, Cantoblanco, Madrid, Spain}

\begin{abstract}{
We discuss the determination of the $f_0(500)$ (or $\sigma$) resonance by  analytic continuation through Pad\'e approximants of the $\pi\pi$-scattering amplitude from the physical region to the pole in the complex energy plane.
The aim  is to analyze the uncertainties of the method, having in view  the fact that  analytic continuation is an ill-posed problem in the sense of Hadamard. Using as input   a class of admissible parameterizations of the scalar-isoscalar $\pi\pi$ partial wave, which satisfy with great accuracy the same set of dispersive constraints,
 we find that the Roy-type integral representations lead to almost identical  pole positions 
 for all of them, while the predictions of the Pad\'e approximants have a larger spread, 
 being  sensitive to features of the input parameterization that are not controlled 
 by the dispersive constraints. Our conservative conclusion 
 is that the  $\sigma$-pole determination by Pad\'e approximants is consistent with the prediction 
 of  Roy-type equations, but has an uncertainty almost a factor two larger. 
 }
\end{abstract}
\maketitle

\section{Introduction}\label{sec:intro}

The determination of a broad resonance like the $I=J=0$ lowest state $f_0(500)$ (known also as $\sigma$) is a notoriously difficult problem. The associated $S$-matrix pole is situated deep in the complex energy plane and, until recent years, the knowledge of $\pi\pi$ scattering at low energies was poor. Therefore,  the extraction of the $\sigma$-resonance parameters was affected by large errors.   For some time, the very existence of this resonance was doubted.
 The  predictions quoted in the current version of PDG \cite{PDG}  still cover a large range, although reduced compared to the previous editions. A  thorough review of the history of the $f_0(500)$ resonance can be found in the recent paper \cite{Pelaez}.

The lack of precision in the early determinations of the $\sigma$ resonance can be related to a great extent to the fact that analytic continuation  is an ill-posed or unstable problem in the Hadamard
sense \cite{Hadamard}, {\em i.e.} arbitrarily small changes in the input data may lead to
indefinitely large variations in the solution. Therefore,  analytic functions which are very close along a finite range  in the complex plane 
may differ arbitrarily much outside it\footnote{Examples of  instability of analytic continuation and its pitfalls in  particle physics have been discussed for the first time in \cite{Ciulli}.}.  In the case of the $\sigma$ resonance, the phenomenon is manifest in a dramatic way because the pole is located far from the physical region  and until recent years no accurate data on $\pi\pi$ scattering at low energies were available.

A major progress was achieved by the use of Chiral Perturbation Theory and dispersion theory,  which  led to a precise theoretical  description of $\pi\pi$ scattering at low energies \cite{CGL, ACGL}. In particular,   Roy equations \cite{Roy}, which fully  exploit analyticity, unitarity and crossing symmetry of the $\pi\pi$ scattering amplitude, are a set of coupled integral equations, whose solutions yield precise values of the partial waves at low energies.   At the same time,  Roy equations provide integral representations  which allow the calculation of the partial waves at complex points in a certain domain  of the first Riemann sheet. The input available along both the right and left cuts by crossing symmetry ensures the stability of the extrapolation to points inside the holomorphy domain. This allowed a first precise determination of the mass and width of $\sigma$ resonance,  reported in \cite{CCL}:
\beq\label{eq:CCL}
m_\sigma = 441\,^{+16}_{-~8}\, {\rm MeV}, \quad \quad \Gamma_\sigma/2= 272\,^{+~9}_{-12.5} \, {\rm MeV}.
\eeq
A similar calculation performed in \cite{BM} confirmed this result, with a somewhat smaller error due to a less conservative estimate of the  uncertainties of the input phase-shifts near 800 MeV.

 Further studies of Roy equations and of their once-subtracted version, known as GKPY equations, were performed in \cite{GKPRY}, including also recent data at low energies from the NA48 experiment \cite{NA48}. The $\sigma$-pole parameters based on GKPY equations read \cite{GKPR}:
\beq\label{eq:GKPRY}
 m_\sigma =457^{+14}_{-13}\, {\rm MeV}, \quad \quad \Gamma_\sigma/2= 279^{+11}_{-~7} \, {\rm MeV}.
\eeq

Recently, a result with a comparable precision  was reported in \cite{Pade}:
\beq\label{eq:Pade}
m_\sigma =453\pm 15 \, {\rm MeV}, \quad \quad \Gamma_\sigma/2= 297 \pm 15\, {\rm MeV}.
\eeq
This result was obtained using a method  based on Pad\'e approximants (PA)  for performing the analytic continuation from the physical region to the pole on the second Riemann sheet \cite{Masjuan:2013jha}.  As starting point for  constructing the  Pad\'e approximants, a specific  parameterization of the scalar isoscalar $\pi\pi$ partial wave at low energies,  given in \cite{GKPRY}, was used. The  parameterization satisfies with great accuracy Roy and GKPY  equations, being therefore a suitable input. However, it might be possible  to find different parameterizations which satisfy to the same extent  the analyticity constraints as the one adopted in \cite{Pade} and may lead to different $\sigma$-pole parameters. Choosing only one parameterization might lead to an underestimate of the true uncertainty of the method.

In the present paper we investigate the uncertainty of the $\sigma$-pole determination by considering a larger class of functions used as starting point in the construction of the Pad\'e approximants.
Our approach is similar to the analysis performed in~\cite{IC}, where it  was shown that the direct analytic continuation of specific parameterizations cannot compete with Chiral Perturbation Theory and Roy equations in the precise determination of the pole associated to the $\sigma$ resonance.
However, while in \cite{IC} the free parameters were fixed by fitting the experimental data on the $\pi\pi$ phase shifts available at low energies, in the present paper we require that the parameterizations satisfy to a great accuracy a set of dispersive constraints. By extrapolating the Pad\'e approximants of these amplitudes to the $\sigma$ pole, as in Ref. \cite{Pade}, we assess in a more realistic way the uncertainty of the pole prediction by this method.  The reliable estimate of the error of the Pad\'e method  will be useful in situations  where the determination of resonance poles is not accessible with Roy or GKPY equations, like in $\pi\pi$ scattering.

The plan of the paper is as follows: in the next section we describe the class of admissible amplitudes used in our study. In Sec. \ref{sec:Roy} we determine the free parameters of the input parameterizations and their statistical uncertainties by imposing the set of dispersive constraints considered in \cite{GKPRY}. In Secs. \ref{sec:poleRoy} and  \ref{sec:polePade}  we calculate the pole parameters of the $\sigma$ resonance for the class of admissible functions, using their contribution to the integral dispersive representations and their Pad\'e approximants, respectively, and discuss also the statistical and systematic errors of the predictions. The last section contains a summary and our conclusions.

\section{Class of admissible amplitudes}\label{sec:class}

We consider the $I=J=0$ $\pi\pi$ partial wave $t_0^0(s)$, which is known to be a real-analytic function in the $s$-complex plane cut for $s\leq 0$ and $s\ge 4 m_\pi^2$ and has a so-called Adler zero at $s\approx m_\pi^2/2$.  On the elastic region of the right cut, which extends to a good approximation up to the $K\bar K$-production threshold, the function $t_0^0(s)$ is expressed as
\begin{equation}\label{eq:t00}
t_0^0(s)=\frac{e^{2i\delta_0^0(s)}-1}{2i\rho(s)},
\end{equation}
where  $\rho(s) = \sqrt{1-4m_{\pi}^2/s}$ and  $\delta_0^0(s)$ is the phase shift.  This relation implies the elastic unitarity relation
 \be \label{eq:unit}
 \mbox{Im} \left[\frac{1}{t_0^0(s+i\epsilon)}\right] = - \rho(s),
\ee
which is valid for $4 m_\pi^2 \le s< 4 m_K^2$.
Therefore, if $t_0^0(s)$ is expressed in general as
\begin{equation}\label{eq:t001}
t_0^0(s)=\frac{1}{\psi(s) - i \rho(s)},
\end{equation}
from (\ref{eq:unit}) it follows that the function $\psi(s)$ is real on the elastic region, where it has the expression
\begin{equation}\label{eq:psi}
\psi(s)=\rho(s)\cot  \delta_0^0(s), \quad 4 m_\pi^2 \le s < 4 m_K^2.
\end{equation}
 The reality property implies also that  $\psi(s)$ is analytic in the $s$-plane cut for $s\le 0$ and $s \ge 4 m_K^2$,  except for the Adler pole  at $s=z_0^2/2$, $z_0\approx m_\pi$.
The parameterization of the partial wave adopted  in \cite{GKPRY} improves the so-called ``effective-range approximation'', which amounts to expanding  $\psi(s)$ in powers of $k^2$, where  $k=1/2\sqrt{s-4 m_\pi^2}$ is the c.m. momentum. It uses the conformal mapping
\beq\label{eq:w}
w(s)=\frac{\sqrt{s}- \sqrt{4 m_K^2-s}}{\sqrt{s}+ \sqrt{4 m_K^2-s}},
\eeq
which maps the $s$ plane cut for $s\le 0$ and $s \ge 4 m_K^2$ onto the unit disc $|w|<1$ in the plane $w\equiv w(s)$.
Then the expansion in powers of $w(s)$ of the form
\bea\label{eq:v1}
&&\psi(s)=\frac{m_\pi^2}{s-\frac{1}{2} z_0^2}\times \\
&&\left\{\frac{z_0^2}{m_\pi \sqrt{s}} + B_0+B_1 w(s)+B_2  w(s)^2+B_3  w(s)^3\right\},\nonumber
\eea
defined as in Eqs. (A1)-(A2) of \cite{GKPRY},  converges in a larger domain and has a better convergence than the simple expansion in powers of $k^2$. We  note that the first term in the second line of (\ref{eq:v1}) removes the singularity of $\rho(s)$ at $s=0$ in the denominator of (\ref{eq:t001}). The parameters $B_n$ of the expansion (\ref{eq:v1}) were taken from the so-called ``CFD parameterization'' of the $\pi\pi$  partial waves, given in Table V of \cite{GKPRY}, which we will describe in detail in Sec. \ref{sec:Roy}.
In Refs. \cite{GKPRY, Pade}, the parameterization  (\ref{eq:v1}) was adopted in the range $4 m_\pi^2\le s \le s_M$, with $\sqrt{s_M}=0.85\,\gev$.  In this paper we shall denote this parameterization of the $S0$ partial wave as $v_1$.

We can construct other parameterizations   by using in the expansion  (\ref{eq:v1}),  instead of the conformal mapping (\ref{eq:w}), a more general mapping $w(s, \alpha)$, defined as:
\beq\label{eq:walf}
w(s, \alpha)=\frac{\sqrt{s}-\alpha \sqrt{4 m_K^2-s}}{\sqrt{s}+\alpha \sqrt{4 m_K^2-s}}.
\eeq
Then we write $\psi(s)$ as:
\bea\label{eq:v2}
&&\psi(s)=\frac{m_\pi^2}{s-\frac{1}{2} z_0^2}\times \\
&&\hspace{-0.45cm}\left\{\frac{z_0^2}{m_\pi \sqrt{s}} + B_0+B_1 w(s, \alpha)+B_2  w(s, \alpha)^2+B_3  w(s, \alpha)^3\right\}.\nonumber
\eea
As discussed in Ref. \cite{IC}, with a proper choice of $\alpha$  the region ($4 m_\pi^2, s_M$) is mapped onto an almost symmetrical range around the origin in the $w$ plane, which ensures a better convergence of the expansion (\ref{eq:v2}) compared to (\ref{eq:v1}). For numerical purposes we will take $\alpha=0.7$. The parameterization defined in this way is denoted as $v_2$.

A somewhat different choice is the so-called Schenk parameterization \cite{Schenk},  adopted in solving Roy equations in \cite{ACGL, CGL, CCL, BM, Regge}. In our notations it corresponds to writing the function $\psi(s)$ entering (\ref{eq:t001}) as
\beq\label{eq:v3}
\psi(s)= \frac{1}{B_0+B_1 k^2+B_2 k^4+B_3 k^6}\,\frac{s-z_0^2}{4 m_\pi^2-z_0^2},
\eeq
where $k$ is the c.m. momentum defined above. The free parameters are the coefficients $B_n$ and $z_0$. We denote this alternative parameterization as $v_3$.

Other parameterizations are obtained using the Chew-Mandelstam procedure \cite{ChMa} of implementing the unitarity relation (\ref{eq:unit}), based on a function which is analytic in the plane cut for $s \ge 4 m_\pi^2$ and has the imaginary part on the cut equal to the factor $\rho(s)$.
For convenience, we  consider the loop function  $\bar{J}(s)$, written as
 \be\label{eq:J}
\bar{J}(s) =\frac{2}{\pi}+\frac{\rho(s)}{\pi} \ln\left[\frac{\rho(s)-1}{\rho(s)+1}\right].
\ee
 It can be checked that this function vanishes at the origin, $\bar{J}(0)=0$, and
\be\label{eq:ImJ}
\mbox{Im}\, \bar{J}(s+i\epsilon)= \rho(s), \quad\quad s\ge 4 m_\pi^2.
\ee
If one defines the function  $\tilde\psi(s)$ by writing:
\be\label{eq:t00J}
t_0^0(s)= \frac{1}{\tilde\psi(s) - \bar{J}(s)},
\ee
the unitarity relation (\ref{eq:unit}) and  the equality (\ref{eq:ImJ}) show that  $\tilde\psi(s)$ is real for  $4 m_\pi^2\le s < 4 m_K^2$, where it is  related to the phase shift $\delta_0^0(s)$ by
\be\label{eq:psitilde}
 \tilde\psi(s)= \rho(s)\, \cot \delta_0^0(s)+ \mbox{Re}\,\bar{J}(s).
\ee
 The reality property implies also that  $\tilde\psi(s)$ is analytic in the $s$-plane cut for $s\le 0$ and $s \ge 4 m_K^2$,  except for the Adler pole  at $s=z_0^2/2$, and can be expanded as
\beq\label{eq:v4}
\tilde\psi(s)= \frac{m_\pi^2}{s-\frac{z_0^2}{2}}[B_0+B_1 w(s)+B_2 w(s)^2+ B_3 w(s)^3],
\eeq
in powers of the  variable (\ref{eq:w}). We remark that the compensating term $\frac{z_0^2}{m_\pi \sqrt{s}}$ appearing in  (\ref{eq:v1}) is no longer
 necessary in (\ref{eq:v4}), since the  function $\bar{J}(s)$ is by definition regular at $s=0$. This parameterization is labeled as $v_4$.

In a similar way, we can expand the function $\tilde \psi(s)$ in powers of the more general
  conformal mapping $w(s, \alpha)$ defined in (\ref{eq:walf}):
\bea\label{eq:v5}
&&\tilde\psi(s)= \frac{m_\pi^2}{s-\frac{1}{2}z_0^2}\\
&&\times\left[B_0+B_1\, w(s,\alpha )+B_2 \,w(s,\alpha )^2+ B_3 \,w(s,\alpha)^3\right].\nonumber
\eea
For numerical purposes we will take $\alpha=0.5$. The corresponding parameterization is denoted as $v_5$.

\begin{table}[t]
\centering
\renewcommand{\arraystretch}{1.3}
\begin{tabular}{l r }
\hline
$t_0^0(s)$ ~~~& ~Equations\\
\hline
~~$v_1$ &  (\ref{eq:t001}),  ~(\ref{eq:v1})\\
~~$v_2$ &  (\ref{eq:t001}), (\ref{eq:v2})\\
~~$v_3$ &  (\ref{eq:t001}), (\ref{eq:v3})\\
~~$v_4$ &  (\ref{eq:t00J}), (\ref{eq:v4})\\
~~$v_5$ &  (\ref{eq:t00J}), (\ref{eq:v5})\\
\hline
\end{tabular}
\caption{Summary of the equations used for the definition of the five parameterizations $v_i$ adopted in the present work. The version $v_1$ is denoted as CFD in \cite{GKPRY}. \label{tab:tab1}}
\end{table}

We summarize in Table~\ref{tab:tab1} the relations used for the definition of the five parameterizations considered in our analysis. As in \cite{Pade}, we have assumed that these expressions are valid on the real axis in the range $4 m_\pi^2\le s\le s_M$, with $\sqrt{s_M}=0.85 \gev$. The free parameters were determined by requiring that the amplitude $t_0^0(s)$ satisfies with great precision the dispersive constraints on the $\pi\pi$ amplitude.  This analysis is presented in the next section.

\vspace{0.3cm}

\section{Dispersive constraints on the admissible parameterizations}\label{sec:Roy}

During the last years, dispersion relations have proved to be a successful tool for describing with high precision  different low-energy hadronic processes (for some examples see \cite{CGL,ACGL,CCL,Buettiker:2003pp,GKPRY,Hoferichter:2015hva}).
 Based on general principles such as Lorentz invariance, causality, unitarity and crossing symmetry,
they allow for a rigorous formalism, which expresses a scattering amplitude at any energy point as a Cauchy integral over the whole energy range.
The dispersive representations  can provide information on the amplitude even at energies where data are poor, in unphysical regions or in the complex plane.
Furthermore, the formalism is model independent, in the sense that the details of the parameterizations used to describe the experimental data become irrelevant once they are used as input in the dispersive integrals.

For $\pi\pi$ scattering crossing symmetry  implies further relations between the left- and right-hand cuts and makes this process specially suited to be analyzed using dispersive techniques. The comprehensive analysis performed in \cite{GKPRY} was based on suitable $\pi\pi$ partial-wave parameterizations obtained from fits to experimental data, but constrained also to satisfy dispersion relations. In particular Forward Dispersion Relations (FDR), once and twice-subtracted Roy equations and two sum rules were imposed as further constraints to the experimental data fits. For completeness, we will summarize next the main characteristics of these dispersion relations.

FDR are fixed-$t$ dispersion relations calculated at the forward or $t=0$ direction~\cite{GKPRY}. They are written in a basis  of $s\leftrightarrow t$ symmetric or antisymmetric amplitudes describing the processes  $\pi^0\pi^0\to \pi^0\pi^0$, $\pi^0\pi^+\to\pi^0\pi^+$ and the amplitude corresponding to the process with isospin one in the $t$-channel.
Standard Roy equations (RE) \cite{Roy} are obtained from the partial wave projection of a twice-subtracted fixed-$t$ dispersion relation,
where the $t$-dependent subtraction terms are determined by $s\leftrightarrow t$ crossing symmetry. This leads to a coupled system of
partial-wave dispersion relations (PWDRs), where the scattering lengths are the only free parameters that appear in the subtraction terms.
Once subtracted Roy or GKPY equations were derived first in~\cite{GKPRY} and, compared to RE, proved to have a slower increase of the uncertainty as the energy grows, making them very well suited for constraining the $\pi\pi$ amplitude in the $f_0(500)$ region. Finally, two sum rules evaluated at threshold were used to constrain the $t$-dependent high-energy Regge behavior in terms of low-energy $P$- and $D$-wave parameters. The first one was constructed in the Pomeron channel whereas the second was defined for the $\rho$ Reggeon exchange.

The determination of the $\pi\pi$ partial waves in \cite{GKPRY}  can be summarized as follow: first a set of simple expressions for each partial wave amplitude are considered to fit separately the available experimental data sets \cite{NA48,pipidata,pipidata2}. Each data set  is checked against FDR and other two dispersive sum rules \cite{GKPRY}.
This leads to an Unconstrained Fit to Data (UFD), where only those experimental data sets compatible within uncertainties with dispersion relations are taken into account.  A recent statistical analysis \cite{Perez:2015pea} has shown that this selection of the experimental data for the UFD violates only slightly the normality requirements of the
residual distributions, which could be fulfilled with tiny modifications of the data selection, leading to almost identical results.

Finally, the parameterizations are used as a starting point for a Constrained Fit to Data (CFD), in which RE and GKPY, as well as FDR, are imposed as additional constraints.
As a result, one obtains a set of consistent parameterizations which, describing the experimental data,
also satisfy the dispersive constraints based on analyticity, crossing symmetry and unitarity, and hence are much more precise and reliable.

\begin{table*}[htdp]
\centering
\renewcommand{\arraystretch}{1.8}
\begin{tabular}{lccccc  }
\hline
&$v_1$&$v_2$&$v_3$&$v_4$&$v_5$\\
\hline
$B_0$ &$7.1 \pm 0.2$ &$10.7\pm 0.5$&$0.22\pm0.01$&$0.36\pm0.02$&$14.9\pm0.5$\\
$B_1$ &$-25.4\pm0.5$ &$-15.3\pm0.3$&$(13.9\pm0.3)$  GeV$^{-2}$&$-59.0\pm0.7$&$-22.2\pm0.9$\\
$B_2$ &$-33.2\pm1.2$ &$-22.5\pm0.8$&$-47.6\pm1.7$ GeV$^{-4}$&$-50.0\pm 1.2$&$-44.9\pm1.6$\\
$B_3$ &$-26.2\pm2.3$ &$-34.0\pm2.9$&$\left(-29.2\pm2.5\right)$ GeV$^{-6}$&$1.04\pm0.01$&$-44.0\pm3.8$\\
$z_0$ &$m_\pi$&$m_\pi$ &0.83 GeV&$m_\pi$&$m_\pi$\\\hline
\end{tabular}
\caption{Values of the parameters of the different $S0$-wave expressions described in Sec. \ref{sec:class}. The choice $v_1$ coincides with the so-called ``CFD'' solution  defined in \cite{GKPRY}.
\label{tab:parameters}}
\end{table*}

 The determination of the $\sigma$ pole reported in \cite{Pade} used the CFD parameterization of the $S0$ partial wave as input  for the analytic extrapolation in the complex plane using Pad\'e approximants. As discussed above, in order to determine the uncertainty of the method in a more exhaustive way, we will use now for the $S0$-wave at low energies the whole class of parameterizations described in Sec. \ref{sec:class}.

\begin{figure*}[htbp]
\begin{center}
\includegraphics[width=0.6\textwidth]{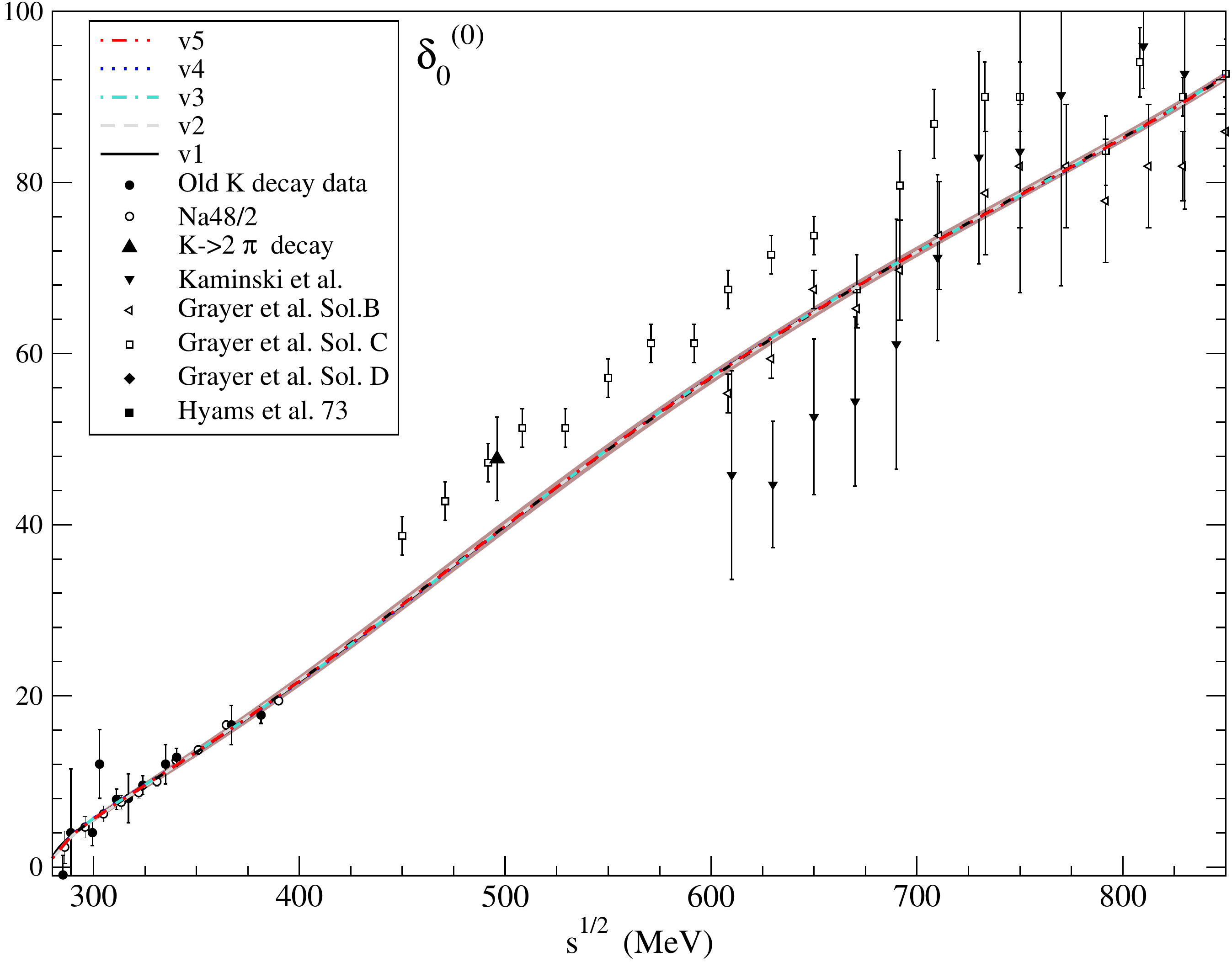}
\caption{$\pi\pi$ S0-wave phase shift for each of the parameterizations discussed in Sec.~\ref{sec:class} in the region between the $\pi\pi$ threshold and $\sqrt{s_M}=0.85 \gev$.
The dark band covers the uncertainties of the $v_1$ (CFD) parameterization. The experimental data points~\cite{pipidata,pipidata2} correspond to those discussed in detail in~\cite{GKPRY}.} \label{fig:S0wave}
\end{center}
\end{figure*}

The easiest way to ensure that the new parameterizations still satisfy the dispersive constraints imposed to the CFD is to determine their free parameters by minimizing the difference between the new $S0$-wave curves  and the CFD one at low energies. In addition,
the errors of the parameters are fixed so that they reproduce the CFD $S0$-wave error bands.
The curves obtained in this way are shown in Fig. \ref{fig:S0wave}, whereas the final parameter values and errors are collected in Table \ref{tab:parameters}.

In order to check whether the new parameterizations are consistent or not with the dispersion relations,
we  use the method applied in \cite{GKPRY}.  
We assume that each of the dispersion relations described above, namely the three FDR, Roy and GKPY equations and the two sum rules, denoted generically as $a$,  
is well satisfied at a point $s_n$, if the difference between its left-hand side and right-hand side ($\Delta_a$) is smaller than its uncertainty ($\delta\Delta_a$), 
which is computed using a Monte Carlo sampling of all the parametrization parameters within 6 standard deviations. 
Thus, if the average discrepancy for a total number of points $N$ verifies
\begin{equation}
\bar{d}_a^2\equiv\frac{1}{N}
\sum_n\left(\frac{\Delta_a(s_n)}{\delta\Delta_a(s_n)}\right)^2\leq1,
\label{avdiscrep}
\end{equation}
we consider that the dispersion relation $a$ is well satisfied within uncertainties.
Following the same convention considered in~\cite{GKPRY}, the values of $s_n^{1/2}$ are taken at intervals of 25 MeV 
between threshold and the maximum energy point defined in \cite{GKPRY} (1420 MeV for the FDR and 1115~MeV for GKPY equations).
The $\bar{d}_a^2$ values obtained for each dispersion relation and parameterization are collected in Table~\ref{tab:CFDdiscrepancies},
and prove that all parameterizations satisfy the dispersion relations equally well. Hence, they are perfectly admissible as input for the Pad\'e approximants in order to extract the $f_0(500)$-pole parameters.
\begin{table}[t]
\centering
\renewcommand{\arraystretch}{1.6}
\begin{tabular}{l ccccc c }
\hline
&&$v_1$&$v_2$&$v_3$&$v_4$&$v_5$\\
\hline
\multirow{3}{*}{FDR}&$\pi^0\pi^0$& 0.53 & 0.64&0.61&0.53&0.50\\
&$\pi^+\pi^0$& 0.43 & 0.43&0.43&0.43&0.43\\
&$I_{t=1}$& 0.22 & 0.23&0.21&0.23&0.23\\
\hline
\multirow{3}{*}{GKPY}&S0& 0.23&0.28&0.26&0.21&0.22\\
&S2& 0.55&0.50&0.53&0.76&0.61\\
&P& 0.11& 0.08&0.10&0.22&0.13\\
\hline
Average&&0.28&0.30&0.29&0.37&0.29\\
\hline
\end{tabular}
\caption{ Average discrepancies $\bar d_a^2$ for each dispersion relation obtained with
different versions of the CFD approach. \label{tab:CFDdiscrepancies}}
\end{table}

\section{Pole determination from GKPY equations}\label{sec:poleRoy}

 The equations denoted as GKPY in \cite{GKPRY} are a coupled system of partial-wave dispersion relations
in which the unphysical left-hand cut is rewritten as a series of integrals over the right-hand cut,
thus expressing the partial waves at any point in the complex plane as integrals along the physical region involving only observable quantities.
The general form of these equations is
\begin{equation}\label{eq:gkpy}
t_J^I(s)=k_J^I+\frac{1}{\pi}\int\limits_{4m_{\pi^2}}^\infty \diff s' \sum\limits_{I'=0}^{2}\sum\limits_{J'=0}^\infty K_{JJ'}^{II'}(s,s')\Im t_{J'}^{I'}(s'),
\end{equation}
where the subtraction terms $k_J^I$ are just linear combinations of the isoscalar and isotensor scattering lengths $a_0^I$,
and the kernels $K_{JJ'}^{II'}$ are composed of a singular Cauchy kernel and a regular remaining piece.

In contrast to  the standard Roy equations \cite{Roy}, GKPY equations are constructed from a once-subtracted fixed-t dispersion relation,
leading to the GKPY constant subtraction terms in \eqref{eq:gkpy}. This produces a smooth error increase at higher energies,
which makes them more suited for the study of the $f_0(500)$ energy region.

An extremely important feature of \eqref{eq:gkpy} is that the Cauchy integrals only require as input the value of the partial waves on the boundary.
Thus, the extrapolation to interior points of two numerically close parameterizations along the boundary provides similar results.
In this way, the extrapolation based on GKPY equation  is ``stable'' in the Hadamard sense,
being  a very suitable framework to perform the analytic continuation into the complex plane of the $\pi\pi$ scattering amplitude.

In Table~\ref{tab:disp-poles}, we collect the $f_0(500)$ pole parameters obtained from the extrapolation to the complex plane of the GKPY equations,
using as input each of the different parameterizations described in Sec.~\ref{sec:class}.
The errors have been computed using a Monte Carlo Gaussian sampling of the
corresponding parameterizations parameters with 7000 samples distributed
within 3 standard deviations. As seen from Table~\ref{tab:disp-poles},
the differences between the pole values are smaller than 1\%,  proving the extremely small dependence of the GKPY equations on the particular parameterization choice.

\begin{table}[t]
\centering
\renewcommand{\arraystretch}{1.5}
\begin{tabular}{rc}
\hline
\toprule
 & $\sqrt{s_\sigma}$ (MeV)  \\ \hline
$v_1$ & $ (457 \pm 14) -i (279\pm 11)$ \\
$v_2$ & $ (456 \pm 13) -i (278\pm 12)  $ \\
$v_3$ & $ (457 \pm 13) -i (279\pm 11)  $ \\
$v_4$ & $ (456 \pm 13) -i (280\pm 11) $ \\
$v_5$ & $ (456 \pm 14) -i (280\pm 11) $ \\\hline
\end{tabular}
\caption{$f_0(500)$ pole positions obtained from the analytic continuation to the complex plane of GKPY equations for each of the different parameterizations detailed in Sec. \ref{sec:class}.}
\label{tab:disp-poles}
\renewcommand{\arraystretch}{1.0}
\end{table}

\section{Pole determination from Pad\'e approximants}\label{sec:polePade}

Pad\'e Theory has been recently considered in Refs.~\cite{Masjuan:2013jha,Pade} as a tool for extracting resonance pole parameters through
the analytic continuation of the elastic scattering amplitude\footnotemark.  
These works investigated the extraction of the mass, width and residue of the resonances using the well-known convergence properties of Pad\'e approximants (PA)~\cite{Baker}.

The approximant denoted as $P^N_M(s,s_0)$ is a rational function with a contact to the function to be approximated of order $N+M+1$ at $s_0$. It is given by the ratio of a polynomial of degree $N$ over another of degree $M$. That means that the expansion of the PA around $s_0$ coincides with the expansion of the function up to the term of $\mathcal{O}((s-s_0)^{N+M})$.

Under quite general conditions~\cite{Baker}, convergence is achieved when $N,M \rightarrow \infty$. In this regard, Pad\'e Theory provides a toolkit~\cite{Masjuan:2013jha,Pade} that allows not only to propose a model-independent method for extracting resonance properties but also to provide a criterion for the evaluation of the theoretical error on the extraction of such resonance parameters. In particular, thanks to the Montessus de Ballore theorem~\cite{Montessus} one is able to unfold the second Riemann sheet of a physical amplitude to search for the position of its resonance poles (if any) 
in the complex plane\footnotemark.

To be more concrete, Montessus' theorem states that if a function $F(s)$ is analytic inside the disk $B_\delta(s_0)\equiv \{s||s-s_0|<\delta \}$ except for a single pole at $s=s_p$, the sequence of one-pole Pad\'{e} approximants $P_1^N(s,s_0)$ at $s_0$,
\begin{equation}\label{PAeq}
P_1^N(s,s_0)=\sum_{k=0}^{N-1}a_k(s-s_0)^k+\frac{a_N(s-s_0)^N}{1-\frac{a_{N+1} }{a_N} (s-s_0)}\, ,
\end{equation}
\noindent
converges to $F(s)$ in any compact subset of the disk excluding the pole $s_p$, i.e,
\begin{equation}\label{th}
\lim_{N\rightarrow \infty} P_1^N (s,s_0)\, =\, F(s)\, .
\end{equation}
As an extra consequence of this theorem, one finds that the PA pole $s_p^{PA}=s_0+\frac{a_N}{a_{N+1}}$ converges to $s_p$ for $N\rightarrow \infty$. The PA coefficients $a_k$ are obtained by matching at a given $s_0$ the expansion of the PA to the expansion of the amplitude $F(s)=\sum_{k=0}^{N+1} a_k (s-s_0)^k+\mathcal{O}((s-s_0)^{N+2})$. If experimental data on the function to be approximated are available along a certain interval, the coefficients $a_k$ can be obtained  by a fit procedure. The use of the derivatives at a fixed point is, however, 
more robust if they are known with enough precision~\cite{Masjuan:2013jha}.

The Montessus' theorem can be generalized to a $P^N_M(s)$ sequence with $M \geq M^*$, where $M^*$ is the number of poles of the original function $F(s)$. In this case, the rest of the poles inside the disk $B_\delta(s_0)$ are unphysical and may be viewed as artifacts that simulate other, more distant singularities, such as branch points produced by unitarity~\cite{IAM-critic}.
The next sequence following the Montessus' theorem would be the approximants with two poles, $P^N_2(s)$. In particular, each element of the $P^N_2(s,s_0)$ sequence around $s_0$ 
is given by (with the definition $a_k=0$ for $k<0$)
\begin{widetext}
\begin{equation}\label{PA2eq}
P_2^N(s,s_0)=\frac{
\sum_{k=0}^N (a_k a_N^2 - a_k a_{N-1} a_{N+1} - a_{k-1} a_N a_{N+1} + a_{k-1} a_{N-1} a_{N+2} + a_{k-2} a_{N+1}^2 - a_{k-2} a_N a_{N+2})(s-s_0)^k}{a_{N}^2- a_{N-1}a_{N+1}+ (a_{N-1}a_{N+2}-a_Na_{N+1} )(s-s_0)  - (a_N a_{N+2}-a_{N+1}^2)(s-s_0)^2}\, ,
\end{equation}
containing two poles located at
\begin{equation}\label{PA2pole}
s_p^{PA} = s_0 + \frac{a_N a_{N+1} - a_{N-1} a_{N+2}}{2(a_{N+1}^2 - a_N a_{N+2})} \pm  \frac{\sqrt{ a_{N-1}^2 a_{N+2}^2- 3 a_N^2 a_{N+1}^2 + 4(a_{N-1}a_{N+1}^3+a_N^3 a_{N+2})- 6 a_{N-1} a_N a_{N+1} a_{N+2} }
}{2(a_{N+1}^2 - a_N a_{N+2})}\,.
\end{equation}
\end{widetext}
One of the poles will converge to $s_p$ for $N\rightarrow \infty$, while the other, together with the polynomial expansion, will simulate other structures.

\footnotetext[2]{Analogous analyses have studied the extraction of the resonance poles through Laurent~\cite{Oller:2015}
and Laurent+Pietarinen expansions~\cite{Svarc:2013}.} 
\footnotetext{Further details such as the theorem proof and other extensions can be found in the book of Baker and Grave-Morris~\cite{Baker}.}

From the above description, it follows that the precise extraction of the function and its higher derivatives at a given point $s_0$ is a necessary condition for the successful application of the method.
In Ref~\cite{Pade}, the amplitude  $t^0_0(s)$  and its derivatives were obtained using the specific  parameterization of the scalar isoscalar $\pi\pi$ phase shift $\delta^0_0(s)$ denoted as $v_1$ in Sec. \ref{sec:class}.  In this section we extend the analysis to all the five parameterizations discussed in Sec. \ref{sec:class}.
For each parameterization, we shall provide the central value of the resonance determination resulting from the central values of the input parameters, and the errors produced by the uncertainties from the input information and the truncation of the PA sequence.
 As in Eqs. (\ref{eq:CCL}) - (\ref{eq:Pade}), we will report the results for the pole position in terms of the mass and width of the resonance, defined by
\beq\label{eq:sp}
\sqrt{s_p}=M-i \Gamma/2.
\eeq
For the truncation error, denoted in \cite{Pade} as ``theoretical error'',   we follow the criterium discussed in Refs.~\cite{Masjuan:2013jha, Pade} and adopt as estimator of the error the difference
\beq\label{eq:error}
\Delta \sqrt{s_{p}^{N}} = \left| \sqrt{s_{p}^{N} } - \sqrt{ s_{p}^{N-1}}\right|
\eeq
for both the mass $M$ and the half-width $\Gamma/2$.

Using the central values and the higher derivatives for each parameterization given in Table~\ref{tab:parameters}, we examined the convergence of the theoretical uncertainty (\ref{eq:error}) of the $P^N_1 (s, s_0 )$ for $N = 0, 1, 2, 3$ and of $P^N_2 (s, s_0 )$ with $N = 0, 1, 2$ for different PA centers $s_0$ in the adopted range of the elastic region (from $\pi\pi$ threshold to 0.85 GeV). The theoretical errors $\Delta \sqrt{s_p^{3}}$ and $\Delta \sqrt{s_p^{2}}$ for the $P^3_1 (s, s_0 )$ and $P^2_2 (s, s_0 )$, respectively, exhibit a minimum at a certain point $s_0$, while the PA sequence is found to break down when $s_0$ approaches either the $\pi\pi$ threshold or the upper end of the considered range.

In order to incorporate the statistical uncertainties coming from the input error bands (Fig.~\ref{fig:S0wave}), we use a Monte Carlo simulation, 
where for a point $s_0$ we generate a set of phase shifts and  derivatives $\{ \delta^{(n)}(s_0)\}$,  with a distribution according to their input errors (assumed to be uncorrelated). The Pad\'e approximants $P^3_1 (s, s_0 )$ and $P^2_2 (s, s_0 )$ are then used to generate a distribution of pole positions.

The procedure is repeated for each $s_0$ and an optimal point, denoted as $s_0^{\textrm{opt}}$, 
is selected in such a way that the quadratic sum of the theoretical and statistical errors of $M$ and $\Gamma/2$ is minimized. The obtained values of $\sqrt{s_0^{\textrm{opt}}}$  range from $470 \mev$ to  $510 \mev$  for $P^3_1(s)$ and from $390 \mev$ to  $510 \mev$ for $P^2_2(s)$. 
The theoretical error induced by the truncation of the PA sequence is of the same order as the uncertainties that stem from the statistical uncertainties of the phase shift. 

After constructing the $P^3_1(s,s_0^{\textrm{opt}})$ and $P^2_2(s,s_0^{\textrm{opt}})$ for each parameterization, 
we extract their pole positions given by the central values of input parameters and collected them in the second column 
in Tables~\ref{tab:P31} and \ref{tab:P22}. The total error (fifth column) is given by the quadratic sum of the truncation theoretical error 
(third column) derived from (\ref{eq:error}) and the statistical error from the Monte Carlo simulation (fourth column), which are assumed Gaussian. 
We computed also the mean value of the statistical Monte Carlo distribution, which turned out to be comparable with the results in the second column
(with deviations below 1~MeV), serving as a cross-check of the normality of the distribution.

\begin{figure*}[htbp]
\begin{center}
\includegraphics[width=0.3\textwidth]{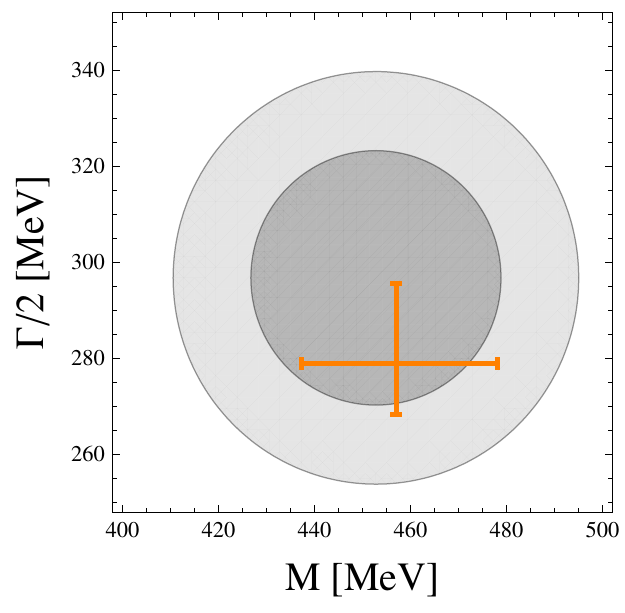}
\includegraphics[width=0.3\textwidth]{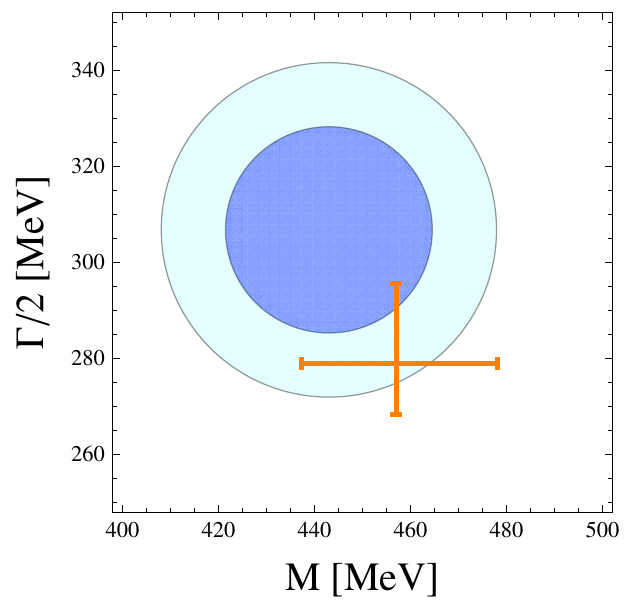}
\includegraphics[width=0.3\textwidth]{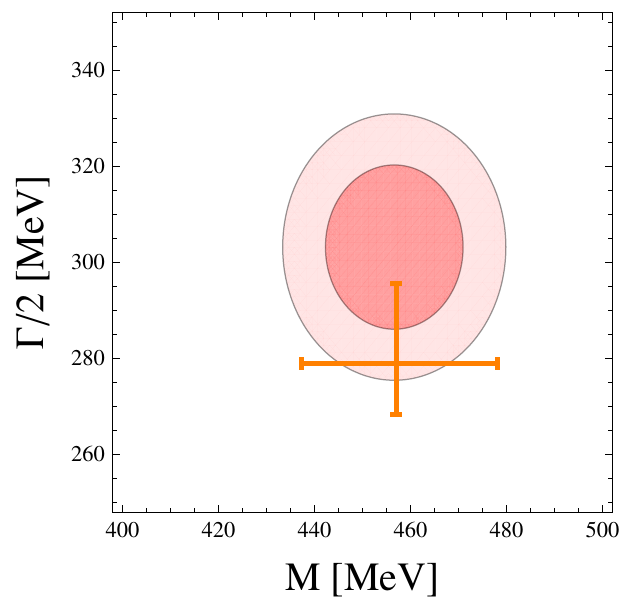}
\includegraphics[width=0.3\textwidth]{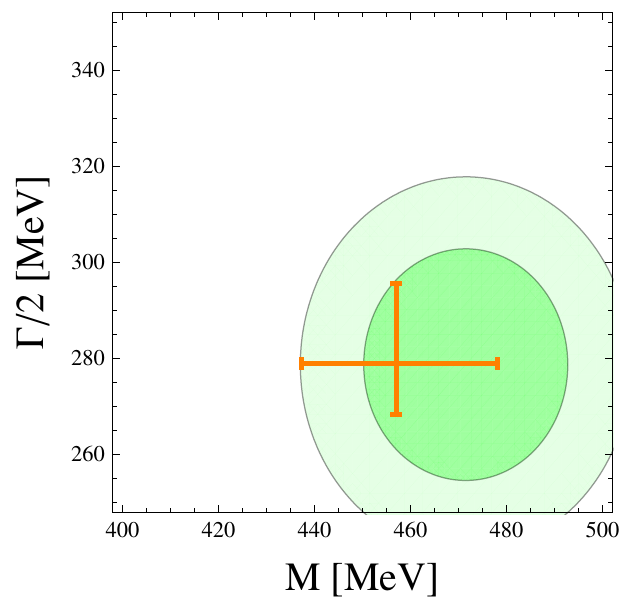}
\includegraphics[width=0.3\textwidth]{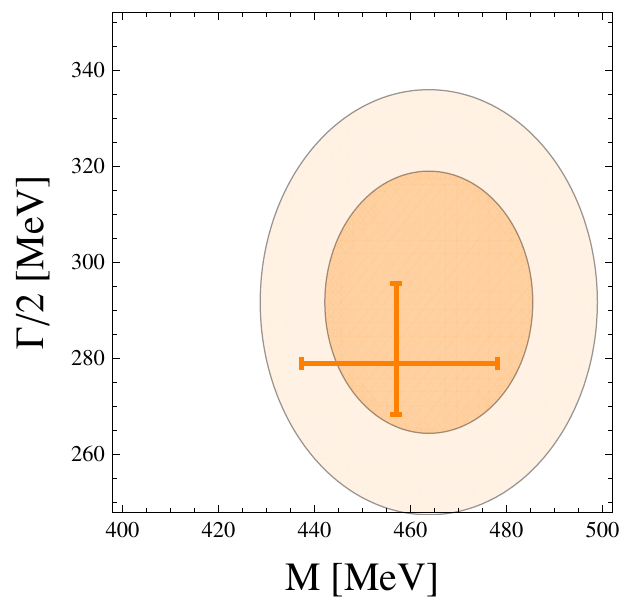}
\includegraphics[width=0.3\textwidth]{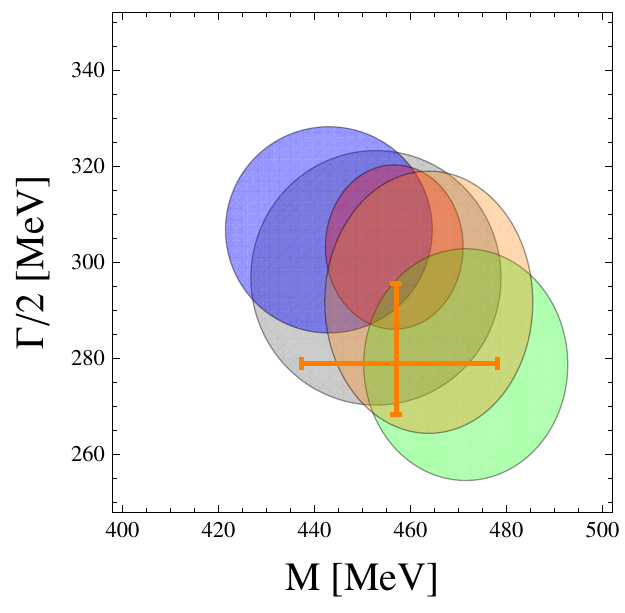}
\caption{Uncertainty regions for the $P^3_1(s,s_0)$ pole determination for each parameterization. Inner ellipses: $68\%$CL; outer ellipses: $95\%$CL. Orange cross: Eq.~(\ref{eq:GKPRY}) shown to guide the eye. Last panel is the overlap of the $68\%$CL ellipses. }
\label{fig:PAN1}
\end{center}
\end{figure*}

\begin{figure*}[htbp]
\begin{center}
\includegraphics[width=0.3\textwidth]{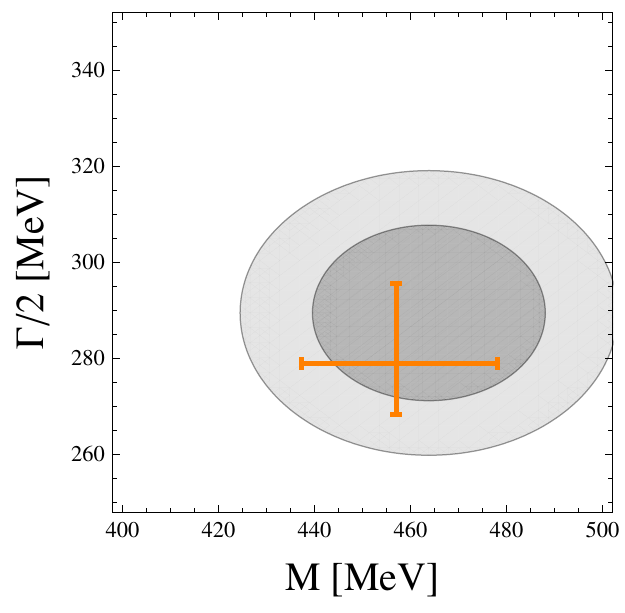}
\includegraphics[width=0.3\textwidth]{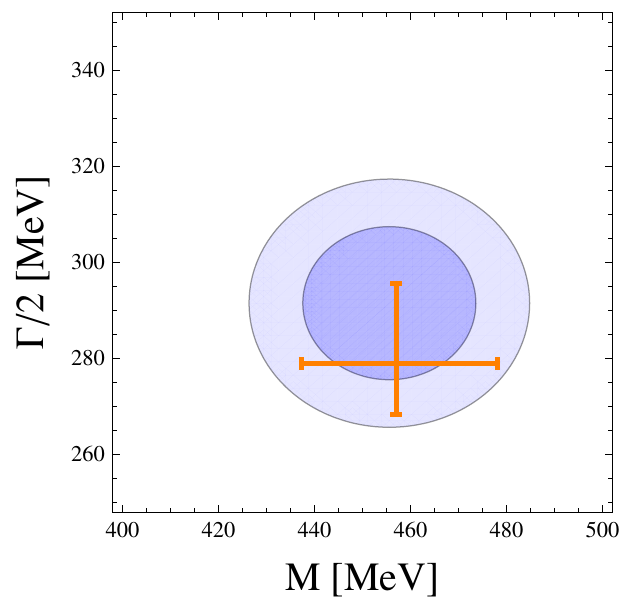}
\includegraphics[width=0.3\textwidth]{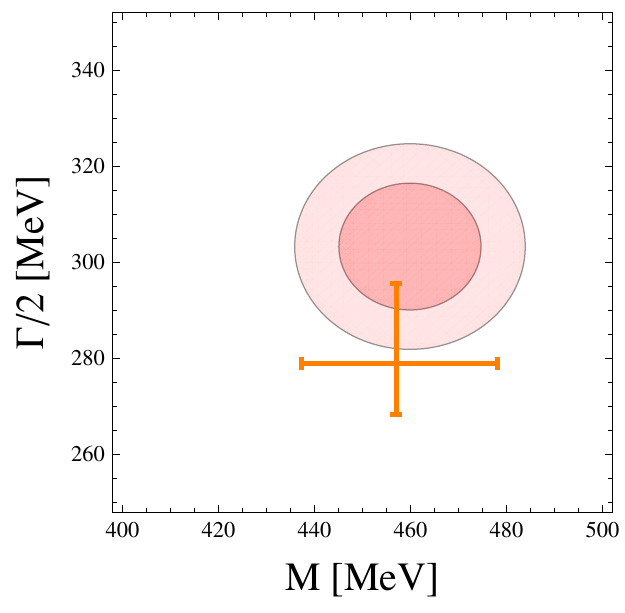}
\includegraphics[width=0.3\textwidth]{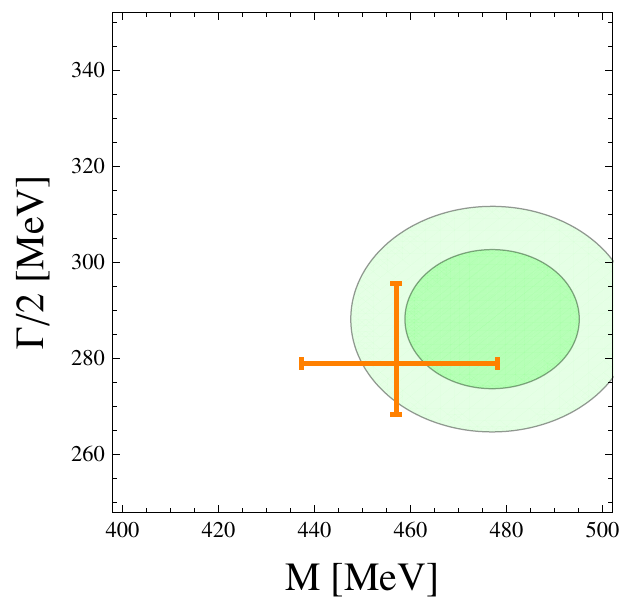}
\includegraphics[width=0.3\textwidth]{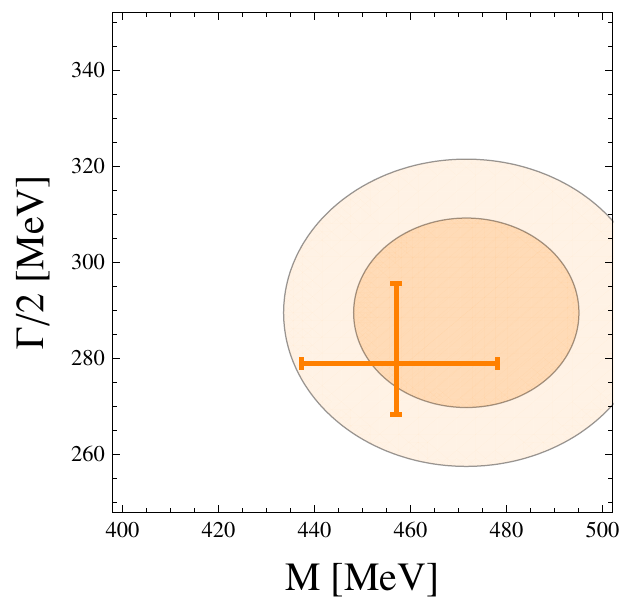}
\includegraphics[width=0.3\textwidth]{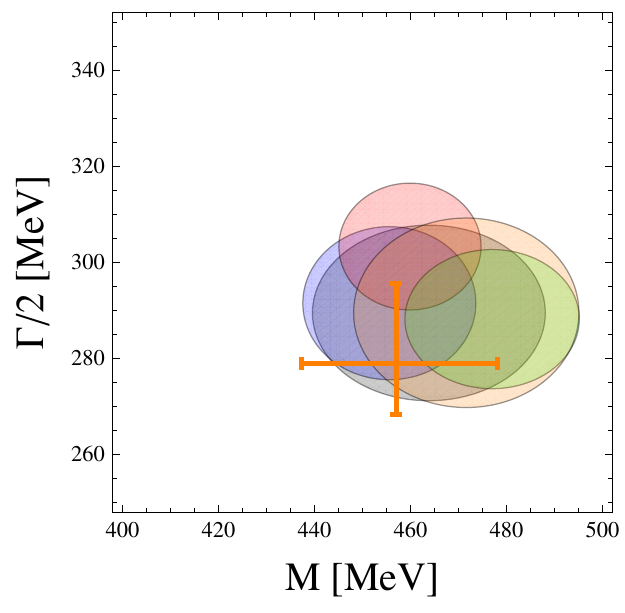}
\caption{Uncertainty regions for the $P^2_2(s,s_0)$ pole determination for each parameterization. Inner ellipses: $68\%$CL; outer ellipses: $95\%$CL. Orange cross: Eq.~(\ref{eq:GKPRY}). Last panel is the overlap of the $68\%$CL ellipses.}
\label{fig:PAN2}
\end{center}
\end{figure*}

Figs. \ref{fig:PAN1} and \ref{fig:PAN2} show the PA results in comparison with a reference value, taken as the prediction (\ref{eq:GKPRY}) of the GKPY equation. The last panels in Figs. \ref{fig:PAN1} and \ref{fig:PAN2} overlap the $68\%$CL predictions for the pole position from the different parameterizations. The spread of the results illustrate the instability of the analytic continuation: the different parameterizations are almost indiscernible in the physical region and satisfy to the same accuracy the dispersive constraints, but lead to rather different pole positions. This shows that choosing a particular parameterization leads to an underestimate of the theoretical uncertainty of the method.

\begin{table*}[htdp]
\caption{Mass and width (in MeV) for the $P^3_1(s)$ approximant.}
\begin{center}
\begin{tabular}{c|cccc}
\hline
& pole & ~~~theo. error& ~~~stat. error & ~~~combined error    \\
\hline
$v_1$ & $M=452.8$ & 14.0 & 10.1 & 17.3 \\
 ($\sqrt{s_0}=500$)& $\Gamma/2=296.8$ & 14.0 & 10.6 & 17.6 \\
\hline					
$v_2$  & $M=443.0$ & 12.3 & 7.2 & 14.3 \\
($\sqrt{s_0}=510$)& $\Gamma/2=306.8$ & 12.3 & 7.2 & 14.2 \\
\hline
$v_3$ & $M=456.6$ & 8.0 & 5.1 & 9.5 \\
 ($\sqrt{s_0}=490$)& $\Gamma/2=303.2$ & 8.0 & 8.0 & 11.3 \\
\hline
$v_4$ & $M=471.5$ & 12.9 & 5.7 & 14.1 \\
($\sqrt{s_0}=470$)& $\Gamma/2=278.7$ & 12.9 & 9.4 & 16.0 \\
\hline
$v_5$ & $M=463.8$ & 10.7 & 9.6 & 14.4 \\
($\sqrt{s_0}=490$)& $\Gamma/2=291.7$ & 10.7 & 14.6 & 18.1 \\
\hline					
\end{tabular}
\end{center}
\label{tab:P31}
\end{table*}%

\begin{table*}[htdp]
\caption{Mass and width (in MeV) for the $P^2_2(s)$ approximant.}
\begin{center}
\begin{tabular}{c|cccc}
\hline
& pole & ~~~theo. error& ~~~stat. error & ~~~combined error   \\
\hline
$v_1$  & $M=463.8$ & 8.8 & 13.5 & 16.1 \\
($\sqrt{s_0}=390$)& $\Gamma/2=290.2$ & 8.8 & 8.4 & 12.1 \\
\hline					
$v_2$ & $M=455.6$ & 8.5 & 8.4 & 11.9 \\
($\sqrt{s_0}=390$) & $\Gamma/2=291.5$ & 8.5 & 6.2 & 10.6 \\
\hline
$v_3$ & $M=459.9$ & 7.2 & 6.6 & 9.8 \\
($\sqrt{s_0}=470$) & $\Gamma/2=303.3$ & 7.2 & 4.9 & 8.8 \\
\hline
$v_4$ & $M=477.0$ & 7.5 & 9.4 & 12.0 \\
($\sqrt{s_0}=510$) & $\Gamma/2=288.2$ & 7.5 & 6.0 & 9.6 \\
\hline
$v_5$  & $M=471.6$ & 10.5 & 11.4 & 15.6 \\
($\sqrt{s_0}=390$)& $\Gamma/2=289.5$ & 10.5 & 7.7 & 13.1 \\
\hline					
\end{tabular}
\end{center}
\label{tab:P22}
\end{table*}%

The problem is to define a more realistic error estimate using these different determinations. The basic principle,  considered also in previous works \cite{IC, GMPY2007}, is to take into account the  spread of the results obtained with indiscernible input in the physical region.  But of course a unique prescription does not exist and some educated guess is necessary. The most conservative strategy would be to take into account the spread of the results seen in the $68\%$CL domains shown in the last panels of Figs. \ref{fig:PAN1} and \ref{fig:PAN2}.  
Taking the extremes of the $1$ standard deviation ranges derived from Tables  \ref{tab:P31} and \ref{tab:P22}, we obtain for the $P^3_1$ and the $P^2_2$ sequences the error intervals: 
\bea
\label{eq:Pcons}
M=(457\pm 28) \mev, \quad \Gamma/2 = (292\pm 29) \mev,\\
\label{eq:Pcons-P22}
M=(466\pm 23) \mev, \quad \Gamma/2 = (294 \pm 18) \mev \, .
\eea

Another possibility would be to include, besides the theoretical and statistical errors given in Tables~\ref{tab:P31} and \ref{tab:P22}, 
an additional uncertainty obtained from the spread of the central predictions of the various parameterizations. This prescription has the advantage of treating separately the various sources of error. The  final central result will be then defined as the average of the individual central predictions  and  the error will be obtained as the quadratic sum of the three independent types of error. 
This procedure leads to slightly smaller errors than in (\ref{eq:Pcons},\ref{eq:Pcons-P22}), namely
\bea
M&=&(457\pm 14 \pm 14 \pm 10) \mev\,=\, (457\pm 22) \mev, 
\nonumber\\
\Gamma/2 &=& (293\pm 14\pm 14\pm 15) \mev \, =\, (293\pm 25) \mev,
\nonumber\\
\label{eq:Pcomb}
\eea
for $P^3_1$ sequence and 
\bea
M&=&(466\pm 11\pm 11\pm 14) \mev\,=\, (466\pm 21) \mev, 
\nonumber\\
\Gamma/2 &=& (296 \pm 8 \pm 11\pm 8) \mev \,=\, (296 \pm 16) \mev,
\nonumber\\
\label{eq:Pcomb-P22}
\eea
for the $P^2_2$ determination. The three errors in the middle part of the identities are the parametrization, the truncation 
and the statistical uncertainties, respectively.

Both approximants give results for the mass and the width compatible with Eq.~(\ref{eq:GKPRY}), 
with larger errors. The $P^2_2$ approximants lead to smaller parametrization, truncation and statistical errors. This may be due to the fact that the expression of the pole of the $P^2_2$ approximant in (\ref{PA2pole}) uses all the input information (all the derivatives at $s_0$),
while the pole of the $P^3_1$ approximant exclusively depends on the ratio of the last two derivatives, which in our case have the largest errors. Likewise,  $P^2_2$  has an extra pole, having this PA sequence a larger flexibility 
and allowing a better approximate description of other singularities  
of the amplitude, like, e.g. the unitarity branch points.

\section{Summary and conclusions}\label{sec:conc}
The problem investigated in this paper is the determination of resonances by using the method of Pad\'e approximants to perform the analytic continuation of the scattering amplitude in the complex energy plane. We considered in particular the determination of the $f_0(500)$ resonance in $\pi\pi$ scattering, which is a notoriously difficult case since the associated pole is situated far from the real axis.  Our analysis  is a continuation of the work reported in \cite{Pade}, having as aim a more detailed investigation of the uncertainties of the pole determination.

The method of  Pad\'e approximants requires as input the value of the amplitude and its higher derivatives at a certain point $s_0$~\cite{Baker}. In \cite{Pade} these values were extracted from a specific parameterization of the $S0$ wave, denoted in \cite{GKPRY} as CFD, which describes with precision the experimental data and obeys a set of dispersive constraints. In order to assess more realistically the uncertainty of the method, we considered a set of admissible parameterizations which satisfy with accuracy the same constraints.
The class described in Sec.~\ref{sec:class} is quite general: all parameterizations are at least as good as  CFD from the point of view of analyticity and unitarity. Just as CFD, they generalize the effective range approximation,  extending its applicability to a larger domain
of the complex energy plane. Moreover, all the parameterizations satisfy with very high precision the dispersive constraints on the cut, being equally good candidates as input for the determination of the lowest resonance. We used this class of functions both in the extrapolation by means of GKPY (Roy-type) equations and in the method of Pad\'e approximants.  From the spread of the pole predictions yielded by these parameterizations, we obtained a better estimate of the uncertainties of the mass and width of the resonance.

Actually, the errors derived by this approach are strictly speaking  only lower bounds on the true uncertainty, since we  restricted the admissible class to a limited set of specific parameterizations\footnote{A parametrization-free approach  (see Ref. \cite{CaCi} for an example) requires more complicated  techniques of functional analysis.}. However, there are no many ways to impose unitarity in the elastic region, and the class of functions considered in the analysis is rather representative from the theoretical point of view. So, we assume that it allows a reasonable estimate of the  uncertainty.

Our final predictions are given in Table \ref{tab:disp-poles} for the extrapolation based on GKPY equations and in  Eqs.~(\ref{eq:Pcons},\ref{eq:Pcons-P22}) and (\ref{eq:Pcomb},\ref{eq:Pcomb-P22}) 
for the extrapolation based on Pad\'e approximants. Our safest estimate, with largest errors, is given by the most conservative approach 
applied to the $P^3_1$ sequence in Eq.~(\ref{eq:Pcons}),
\beq\label{eq:final}
M =(457\pm 28) \mev, \quad \Gamma/2 = (292\pm 29) \mev\,. 
\eeq

The Roy-type integral representation leads to almost identical predictions for all parameterizations, therefore the error given in the original GKPY  result (\ref{eq:GKPRY}) is not modified. On the other hand, the outcome of the direct analytic continuation through Pad\'e approximants has  a larger spread. This spread defines a new source of error, related to the instability of analytic continuation of functions almost indiscernible in the physical region.   The origin of the larger errors lies in the  higher derivatives of the function, used as input in the Pad\'e approximants.  The   higher derivatives are not controlled by the constraints of unitarity and analyticity, and  can be quite different for functions which satisfy with the same accuracy the dispersive constraints.  These differences play a crucial role in the extrapolation to the pole in the complex plane.

Compared with the values given in Table \ref{tab:disp-poles}, the most conservative result given in (\ref{eq:final}) has an error larger by a factor of about two. On the other hand, it is important to emphasize that the intervals provided by the Pad\'e approximants in \eqref{eq:Pcons}-\eqref{eq:Pcomb-P22} are perfectly consistent within errors with the precise values obtained with the Roy-type integral representation. 
This gives confidence in the method. 

In this work we have not further explored other possible approaches to reduce the Pad\'e uncertainties. The aim of this work was to restudy the outcome of~\cite{Pade} which relied on the value of the phase-shift $\delta(s_0)$ and its first four derivatives $\delta^{(n)}(s_0)$. 
Considering additional derivatives would allow us to reach higher orders in the PA sequence and to decrease the truncation error. However, the statistical uncertainty is expected to increase. The study of the global error may tell if this can lead to a neat improvement. The incorporation of the precise knowledge on the scattering lengths \cite{CGL, GKPRY} can be also used to stabilize the Pad\'e approximants and their pole prediction. Other approach that can be explored in future works consists on using directly the outcome Re $t^0_0(s)$ from the GKPY dispersion relation for the construction of the PA, which is expected to decrease the impact of the parametrization ambiguity. Our conclusions can be applied to the methods discussed in Refs.~\cite{Oller:2015,Svarc:2013} as well.

We recall finally that for constructing the Pad\'e approximants we resorted as in~\cite{Pade} to the precise information on the $S0$ partial wave provided by the dispersive constraints. Actually, in the case of $\pi\pi$ scattering the use of Pad\'e approximants might seem unnecessary, since the straightforward analytic continuation of Roy or Roy-type equations offers the most precise way of finding the low-energy resonances~\cite{CCL, BM, GKPRY}. However, by identifying the origin of the instability of the analytic continuation and by assessing in a more exhaustive way the uncertainties of Pad\'e approximants, the investigation performed in this work provides a common methodology for the analysis of further reactions and observables, for which dispersive techniques cannot be applied but which can be addressed by means of the Pad\'e approximant method.

\subsection*{Acknowledgments}
The work of IC was supported by  UEFISCDI under Contract Idei-PCE No 121/2011. 
The work of PM and JRE was supported by the Deutsche Forschungsgemeinschaft DFG through the Collaborative Research Center 
``The Low-Energy Frontier of the Standard Model" (SFB 1044) and ``Subnuclear Structure of Matter'' 
(SFB/TR 16) respectively.  
The work of JJSC was supported by the Spanish Ministry MINECO under grant FPA2013-44773-P, 
Consolider-Ingenio CPAN CSD2007-00042 
and the Centro de Excelencia Severo Ochoa Programme under grant SEV-2012-0249.

\end{document}